# Secure Implementation of a Vehicular Application for the Symbian Platform

*F. Martín-Fernández, C. Caballero-Gil, P. Caballero-Gil, J. Molina-Gil*

University of La Laguna, Tenerife,
e-mail(s): fmartinfdez@gmail.com, {ccabgil, pcaballe, jmmolina}@ull.es
Spain

**Abstract:** A vehicular ad-hoc network is a type of communication network where the nodes are vehicles. Their aim is to manage traffic efficiently in order to prevent unwanted events such as accidents, jams, etc. The research on this type of networks assumes the use of additional infrastructure installed on the roads, and special devices installed in vehicles. In this paper we propose a novel solution for the deployment of vehicular networks, which uses only existing smartphones with technologies such as Wi-Fi, GPS and Bluetooth. This approach will save the costs of road infrastructure deployment and extra vehicles devices, making it possible to deploy the first real-world vehicular ad-hoc network with a low cost. In particular, the developed implementation of the proposal for the Symbian platform is here examined in full detail.

## 1. INTRODUCTION

Road safety has become one of the major concerns of recent times. Preventing accidents on the roads is one of the common objectives of all nations. In particular, road safety is a fundamental topic because the consequences of traffic accidents are losses of lives. Additionally, traffic jams have a negative effect on the economy and the environment, so the conclusion is that roads involve a complex problem that requires an urgent solution.

One measure that has emerged in the literature recently is the introduction of autonomous vehicular networks or VANETs (Vehicular Ad-hoc NETworks). In these networks, vehicles communicate with each other to prevent adverse circumstances on the roads and to achieve a more efficient traffic management. However, the actual implementation of these networks such as has been proposed by most researchers would involve an economic effort by governments and users, what so far has been a serious impediment.

This paper provides the implementation of a tool for driver assistance, called VAiPho (VANET in Phones), which creates the first real vehicular network using only smartphones to make possible the current deployment of VANETs. This tool is the result of the implementation of the patent presented by the ULL (University of La Laguna) in 2010 [1].

Specifically, the development of VAiPho is here proposed for smartphones with Symbian operating system, which provides not only the automatic detection of traffic events in real-time, but also secure communications between vehicles in order to issue warnings about such events. Regarding secure communications, this paper includes the implementation of a strong authentication scheme based on distributed trust among users, as well as several cryptographic algorithms to verify the accuracy of node identities. Moreover, event detection is performed optimally and efficiently in order to ensure where and when a certain traffic condition happens. Also, a data aggregation scheme is included to allow verifying with other users that an event is real, and thus preventing erroneous or malicious messages.

This paper is organized as follows. Section 2 includes a brief description of related research, whereas Sections 3 and 4 show the general structure of VAiPho, giving details of its modules and interfaces. Finally, conclusions close the paper.

## 2. RELATED RESEARCH

A VANET can be seen as an extension of a Mobile Ad-hoc NETwork (MANET), with a technology that uses moving vehicles as nodes of the network. This makes all involved vehicles would act at the same time as routers and as nodes capable of communicating with other nodes and with the traffic system.

The history of the use of infrared and radio communications among vehicles and infrastructure on the roads is strongly linked to the evolution of the Intelligent Transportation System (ITS). The use of communications in vehicular environments for improving road safety and efficiency was first addressed in the 1939 World's Fair where General Motors presented a vision of a future city called "Futurama". In the late 60's, systems based on radio communications began to be developed on roads, showing their benefits. Since then, the field has progressed significantly, but the focus of the research has been constantly changing over time. At first it was focused into systems that guided the driver to follow directions, but then it changed to put more emphasis on toll systems. Afterwards, the new goal is research on autonomous driving and the field has become more popular in scientific research. However, the central objective of the research has never been altered because it has been always the improvement of road safety and efficiency.

Research and development in these areas are mainly conducted in the United States, but Japan and Europe were not far behind and also made important contributions in this field.

The first proposals for vehicular ad-hoc networks appeared in the decade of 1970, so this year may be considered the starting point of the conceptual history of VANETs. The work [17], published in such a year, proposed an electronic guided tour denoted SRGS (Self- organizing Route Guidance System). For implementing a prototype of SRGS, antennas were installed at the intersections of roads and under the rear of vehicles so that the corresponding communication system operated at 170 kHz, and the data transmission rate was around the 2000 bits per second. According to what the work [4] stated, the efforts made by Rosen to make its system a reality was cut short due to the high cost involved in installing all necessary infrastructure on roads.

In Japan, a project to control car traffic called CACS (Comprehensive Automobile Traffic Control System) was carried out between 1973 and 1979 by the Agency of Industrial Science and Technology, Ministry of International Trade and Industry. The objectives of this project, explained in [10], are still valid after more than 30 years:

- Reduction of traffic congestions.
- Reduction of gases produced by cars because of traffic congestion.
- Accident prevention.
- Improvement of the public and social role of the automobile.
- Provision of information and priority to emergency vehicles.

CACS project conducted a pilot proof-of-concept with 98 units of roadside equipment and 330 test vehicles, as reported by [15]. For the communication between road stations and vehicles, pump road and ferrite antennas in vehicles were used, so that the transmission rate rose to 4.8 Kbps.

Afterwards, in Europe, the PROMETHEUS (PROgraM for European Traffic with Highest Efficiency and Unprecedented Safety) started in 1986 and was launched in 1988 to stimulate research and development in the field of information technology and mobile communications between vehicles and road. This program was developed by 19 European countries and supported by the European Commission, as described [23] and [24]. The publication [6] explains the organization of PROMETHEUS, which is structured into a series of subprograms:

- PRO-CAR: driver assistance by electronic systems.
- PRO-NET: vehicle-to-vehicle communications.
- PRO-ROAD: vehicle-to-environment communications.

The report [2] about the program can be still considered acceptable today in many aspects. It discussed the communication needs based on typical scenarios, such as for example, a simple lane change maneuver. Assuming a periodic dissemination strategy and accuracy of collision distance, the report showed that with a conservative estimate of the rate of periodic transmission of status messages between vehicles, each vehicle should perform 20 of these transmissions per second, so the introduction of prediction methods was proposed to reduce the number of required transmissions. Thus, in order to achieve a lower communication rate, the focus was put on systems operating in the frequency band 60 GHz [5].

The interest of communication between vehicles continued in Japan and the United States. [10] cites two technical reports published in 1986 and 1988 respectively, which helps to implement experimental results from vehicle to vehicle communication. Meanwhile, in the U.S., as indicated [20], the main line of research was related to the autonomous driving of the car. The works [19] and [12] were the first to define some of the basics of automation in the road. Sachs [18] presented the requirements and benefits necessary to carry out both vehicle-to-vehicle communications and communications between the vehicle and infrastructures on the road.

Jurgen [19] described the status of existing related projects at that time in the U.S., Japan and Europe as well as a vision of the role of the necessary systems to enable intelligent vehicles on the roads.

The unchanging key concept in the history of VANETs is the financing of the system. From the beginning, entrepreneurs have wondered how the VANET concept could be exploited and maintained. Different possibilities have been proposed: communication-based tolls, congestion fees, etc. Although these and other techniques have been proposed as possible solutions, skepticism persists. The work [9] described this problem as "The chicken-and-egg paradox ". On the one hand, the automotive and electronics industries doubt whether the public infrastructure for ITS will ever materialize. On the other hand, potential users of the system and governments doubt whether the ITS technology could ever offer practical solutions to the current road traffic problems.

In the second half of the 1990s there were significant changes related to this field paradigm. In San Diego in 1997 within a demonstration organized by the California Partners for Advanced Transit and Highways (PATH), many advances occurred in automated highway driving. It was not the only conference that showed something like this. In the city of Tsukuba in Japan, in 2000, something similar was presented during a demonstration of the project called ASV (Advanced Safety Vehicle). Europe would not be less, so the EU project PROMOTE CHAUFFEUR also presented related results.

Then, the target suddenly changed to try to reach autonomous driving through cooperation in order to try to create a cooperative system for driving assistance. From 1998 to 2005, the United States, within the heart of the Intelligent Vehicle Initiative (IVI), as was explained in [11], research was developed to seek road safety through active cooperation. Also two European projects, FleetNET and CarTalk, as described by [3], investigated the technologies and applications that could provide assistance to drivers cooperatively. In Japan, the ASV Project Phase 3 acknowledged the key role that communication should take for the vehicle driving assistance in a cooperative way.

The huge advances in technology and standardization since the mid 90's years have considerably affected VANET idea. In 1999 there was a very important turn when the Federal Communications Commission in the United States allocated 75 MHz of bandwidth in the 5.9 GHz band of the total to DSRC (Dedicated Short-Range Communication). This term denotes a set of protocols and standards for short-range wireless communication between vehicles and infrastructure, and aims to provide technological neutrality.

A year later, the ASTM (American Society for Testing and Materials) established a working group to create the conditions for the new DSRC standard. In 2001, the Standards Committee 17.51 of the ASTM selected IEEE 802.11a as the underlying radio technology for DSRC. The corresponding standard was finally published in the year 2002 (ASTM E2213-02 2003) and revised in 2003 (ASTM E2213-03 2003). The pressure to use the assigned channels and the availability of both IEEE 802.11a and technology standards increased significantly the research and development in this field, and the interest of the community of mobile networks in the area of vehicular networks was revitalized.

In 2004 the IEEE association began to work on the IEEE Standard 802.11p for wireless access in vehicular environments known as WAVE (Wireless Access in Vehicular Environments), and based on the ASTM standard, as described in [8]. The security in vehicular communications was studied in the project called VSC (Vehicle Safety Communications) supported by the association CAMP (Crash Avoidance Metrics Partnership), the U.S. Federal Highway Administration and the U.S. National Highway Traffic Safety Administration. The project investigated the DSRC technology between 2002 and 2004, concluding that the approach based on IEEE 802.11a would be able to withstand most security applications that the VSC project had proved. The final report, published in 2006, discussed some weaknesses of the standard, such as low communication latency, high availability of the radio channel or general topics related to the channel capacity. In 2004, the first international meeting organized by the ACM on vehicular ad-hoc networks coined the term VANET.

One of the main requirements of any VANET is that users can authenticate it safely. Data privacy is also essential to use this technology without any fear of being tracked or accessing any private information. Many experts have proposed different solutions to these security problems. For instance, methods based on sending keys [16] or on anonymity [22] try to solve this paradigm. The detection of fraudulent users is another aspect to consider in VANETs. Solutions like [7] have been proposed for this problem.

Currently many research projects on VANETs have being carried out, including the study of applications to different areas, as shown in (Technology Solutions to Improve Transportation Safety and Reduce Congestion. 2008) and Maile [13]. However, all the proposed applications have the same main handicap, which is the need to install special equipment on the roads and special devices in vehicles because this involves a high cost, as it means more expensive cars and spending public money to deploy road infrastructures. The proposal VAiPho analyzed here avoids this problem, because it proposes an innovative solution using existing mobile devices, with Bluetooth, GPS and Wi-Fi. Later sections explain in detail this proposal, which will allow the creation of the first real-world VANET, with no additional cost.

### 3. MODULES OF THE PROPOSAL

For the implementation of VAiPho, in this work the chosen operating system was Symbian, of the Nokia Company. One factor in this choice was that its terminals were, with a big difference, the most popular mobile phones, as shown in Figure 1.

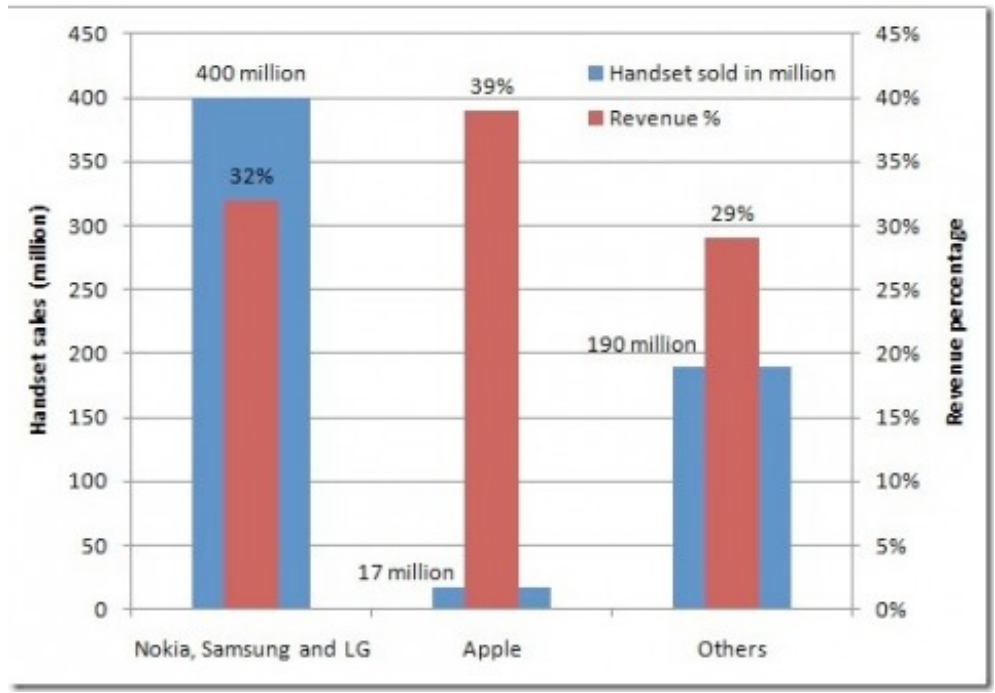

*Figure 1. Mobile phones sales in 2010*

This work was developed after the VAiPho application had a well-defined structure, explained in [1]. The intention was that VAiPho for Symbian were an application designed and implemented modularly in order to allow greater flexibility for future improvements.

### 3.1. Network Connections

The first consideration was the need to connect the device to a network. Furthermore, the network should be ad-hoc and created spontaneously from the application, and the first terminal using VAiPho should create the ad-hoc network because the network would not exist previously. In this way, successive users who use VAiPho will connect with the first user that created the network and will not have to create any new network. Therefore, if the first user, the creator of the network, leaves the network and other users are connected, the ad-hoc network will continue existing since each user has all the necessary information to continue transmitting data through it.

The network to be created spontaneously should have a common name for all terminals can recognize it. It was decided that the name was the name of the application, vaipho. Thus, the application can recognize if there is an ad-hoc network named vaipho, and then, either it connects to the existing one, or create it. This is enough to allow the application to properly connect to the ad-hoc network. Several connection times have been measured depending on whether it was necessary to create it or not, and regardless of whether the application was running on the mobile terminal or in the simulator development environment. As shown in Table 1, the computer simulator requires for the connections a much shorter time than the mobile terminal.

*Table 1. Times for the network connection*

| Time in seconds | Network already created | Network not yet created |
|---|---|---|
| Simulator | < 1 second | Between 1 and 3 |
| Smartphone | Between 15 and 20 | Between 40 and 60 |

### 3.2. Client-Server Module

The next step was the creation of a communication module according to the client–server architecture. In order to do this, the application must act as both the client and the server. This peculiarity is done by multithreading, creating a thread to handle the client and another thread to handle the server. Thus, the application could send packets to the network via the client and be continually listening to receive packets from the server.

An appropriate configuration was selected to perform sending and receiving on the network. It was decided the use of a transport layer protocol based on the exchange of data packets, known as UDP (User Datagram Protocol). This protocol uses a specific port to send and to listen. It allows sending UDP datagrams through the network without previously established connections, since the same datagram routing incorporates enough information in its header. There is no flow control confirmation, so that packets can overtake each other, and it is not known whether the packet has arrived properly because there is no receipt of delivery confirmation. Its main use is for protocols where connections and disconnections are fast, but do not require any information on successful transmissions. This is the case of many audio and video transmissions in real time with strict delay requirements.

It was established that every node should be identified by a variable PSEUdonym (PSEU), which is used to uniquely identify each user as it serves as an alias to be released to other nodes.

The client module has three distinct functions:

1.    Answer: This function is to answer to a given node of the network that sent a datagram. The node's server deals this packet. These answers are very common both in the authentication and the data aggregation phases. In addition, to verify whether several answers are from the same user, since in the UDP protocol does not provide certainty on the origin of a packet, an algorithm was created to detect whether a message arrived or not. This is achieved through the submission of additional answers, i.e. an automatic answer is sent whenever a message is received. Knowing this, when sending a message a node can ensure whether it arrived the destination by getting the expected answer. Thus, it is necessary to wait some time when the expected response has not been received, before suspecting that other user has not receives the message and sending it again. This procedure involves posting messages if an answer is not received. To solve this issue and avoid that this loop is infinite, it was established a maximum limit of 3 re-sending of any message. If this limit is exceeded, the program aborts the expected-answer phase. This mechanism is used for authentication, as will be shown below.

2.    Beacon: The client must also be able to send beacons periodically to the network in a broadcast mode in order to inform other users of the network that it has not yet been authenticated. Every node must be active and ready to start a new authentication cycle all the time. This submission should be constant but a random wait time between shipment and delivery is needed to eliminate any possibility that the signal of a particular user can be monitored. In order to send this message in a broadcast mode, it will be enough to specify it in a parameter.

3.    Changing Pseudonym: The user's pseudonym is updated from time to time. Therefore, the client must be able to notify this new pseudonym to the devices that are already authenticated with this user. Every few beacons sent with a node's pseudonym, it has to be changed in order to prevent tracking, so that the new pseudonym is notified to the authenticated devices. In addition, every node makes all the necessary changes in its database.

The client module is responsible for sending the relevant data to the network. This can be done to a specific address of the network or otherwise, to the entire network in a broadcast mode, as shown in Table 2.

*Table 2. Types of VAiPho messages*

| Message Header | Specification | Message Format |
|---|---|---|
| 01 | Beacon Annunciation | 01,PSEU,DATE,CTA* |

| 01 | Change PSEU | 01,PSEU,DATE,00, NEWPSEU,CTA2* |
|----|-------------|--------------------------------|
| D1 .. D5 | Discovery | COD,PSEU,INFO* |
| Z2 .. Z4 | Zero Knowledge Proof | COD,PSEU,INFO* |
| E1 .. E6 | Exchange | COD,PSEU,INFO* |
| T1 | Traffic Jam Event | COD,PSEU,INFOT1* |
| P1 | Publicity Event | COD,PSEU,INFO* |
| P2 | Parking Event | COD,PSEU,INFO* |

* PSEU: User pseudonym.
DATE: Exact date of message sending.
CTA: Frame that specifies the format: EK1 (ID1:KUid1:TimeStamp).
NEWPSEU: New user pseudonym.
CTA2: Frame that specifies the format: EK1 (00:TimeStamp:newPseu).
COD: Message header: D1 .. D5 | Z2 .. Z4 | E1 .. E6 | T1 | P1 | P2.
INFO: Message data.
INFOT1: Message data of T1 Types: I, F & A.

The server module is responsible for receiving all packets that arrive directly or indirectly to the application via the network.

### 3.3. Authentication Module

Once the application is fully functional with respect to connections within the ad-hoc network, the next phase is the authentication of users, which is the basic tool of the application. In order to use VAiPho, users need demonstrating that they are trustful, and not fraudulent. This is achieved through an authentication protocol, which is always executed between two users who fulfil the condition that both of them have a common acquaintance. The proposed algorithm uses a message passing to verify the identity of another user in order to be able to discern whether it is legitimate or not. The cryptographic protocol consists in a mutual authentication algorithm based on an interactive challenge-response protocol, which is a zero-knowledge proof.

Note that it is possible to authenticate only one user at a time. At the beginning of this authentication process, one of the users, here called Alice, receives one of the beacons (01) sent in broadcast mode by the other user, here called Bob. From that moment the process that will end with the authentication of both users begins.

Alice then sends the list of IDs of her acquaintances, in order to allow Bob discovering whether there is some common acquaintance between Alice and Bob, what will provide user authentication reliability.

Also, Bob receives the D3 message sent by Alice, and tries to check whether the hash sent by Alice in the D1 message corresponds with the resulting hash obtained from data sent by Alice in the D3 message.

Once Bob sends the D4 message with the node in common between the two users, Alice generates a random graph from the public key of the node in common, which is used to create an isomorphic graph to be sent back to Bob in order to find out whether it really knows the key of the node in common.

In particular, the random graph is created from the public key in binary by substituting some 0s by 1s arbitrarily. Note that this information is not useful to nodes that have not the common public key, because it is necessary to know such a public key to generate the starting graph. Its adjacency matrix has the property of having two 1s per row and column, which is related to the existence of a Hamiltonian circuit in the graph. Another relevant consideration in the implementation was to keep the public key length in 15 bits.

*Example:*
Public key: 12869
Public key (in binary): 0 1 1 0 0 1 0 0 1 0 0 0 1 0 1
Starting Graph:
0 0 1 1 0 0
0 0 1 0 0 1
1 1 0 0 0 0
1 0 0 0 1 0
0 0 0 1 0 1
0 1 0 0 1 0
Randomly transformed public key: 0 1 1 1 1 1 0 1 1 0 1 0 1 1 1
Graph according to the transformed public key:
0 0 1 1 1 1
0 0 1 0 1 1
1 1 0 0 1 0
1 0 0 0 1 1
1 1 1 1 0 1
1 1 0 1 1 0
With this graph, the node creates an isomorphic graph, by changing the positions of rows at random according to a random vector.

Random vector: [6, 5, 3, 4, 2, 1]
Isomorphic Graph:
1 1 0 1 1 0
1 1 1 1 0 1
1 1 0 0 1 0
1 0 0 0 1 1
0 0 1 0 1 1
0 0 1 1 1 1

After receiving the isomorphic graph in the D5 message, Bob generates a random challenge (a binary number, 0 or 1), and sends it to Alice in the message Z2. Then, depending on the received question, Alice sends a different piece of information. Thus, if the question is type 0, Alice sends in the Z3 message, information to show Bob the chosen graph isomorphism. However, if the question is of type 1, in the Z3 message, Alice sends information to Bob so that this can verify that the isomorphic graph has a Hamiltonian circuit.

Z3 message received by Bob contains the necessary information to verify the correct response to the challenge. If it works, Bob sends a new random question to Alice in Z4 message, and as before, Alice generates the appropriate answer, E1 message, with the necessary data to answer the question. When Bob receives the packet E1, after verifying that the received information is correct, it decrypts the data with the public key of Alice, adds this user to its database and sends the E2 message, containing its public key. Bob is then added to A's database.

When Bob receives E3 message, it marks Alice as authenticated and sends Alice the E4 message with data from Bob. Then, it marks Bob as authenticated. In order to send an E5 message to Bob, Alice checks that it has successfully received the data. In addition, once authenticated, it sends the events about traffic conditions stored in its database. When Bob receives the E5 message to notify Alice that it has received the messages, Alice sends a final E6 message to finish the authentication process. At this moment, Bob sends the events stored in its database.

The time a node takes to authenticate to another node has been measured taking into account different considerations. Also, again we compare the time taken by the simulator and the time taken by the actual mobile terminal, as shown in Table 3, and distinguish among different numbers of users and whether the GPS signal is available or not.

*Table 3. VAiPho authentication times.*

| Time in seconds | 2 Users | | 3 Users | | 4 or more Users | |
|---|---|---|---|---|---|---|
| | GPS | Not GPS | GPS | Not GPS | GPS | Not GPS |
| Simulator | < 0.5 | < 1 | < 1 | < 2 | < 1.5 | < 2 |
| SmartPhone | 3 - 5 | 5 -7 | 7 - 10 | 12 - 15 | 7 - 10 | 12-20 |

### 3.4. Event Detection

Once solved communication and authentication in VAiPho, the next problem is the automatic detection of events by the application. This aspect is considered in three different types of traffic conditions: free parking and location of parked vehicle and traffic jam, and a type of special event that is geo-located advertising.

•    Free parking: This event indicates to the user a possible free parking nearby. The free parking space is announced only as possible, because the application does not have the power to book any parking space so if someone else arrives before, VAiPho cannot detect it. Thus, this event has an associated parameter that indicates the validity time. The duration of the event can be configured by the user of the application and is added to the time of detection of the event to calculate the expiration time. A very simple concept is taken into account for the automatic detection of this event: a parking space is released when a vehicle leaves it. Thus, the user releases the parking space automatically when it starts the vehicle as it automatically launches VAiPho. This type of event is only visible when the user chooses the option to do so, i.e. when it presses the button to search for parking. When the application is launched, if it has GPS satellite coverage, it automatically detects the free parking event, inserts it into its table events and notifies it to the authenticated users through a broadcast that announces its current GPS coordinates.

•    Vehicle Parking: Another event that VAiPho can handle is based on storing the information needed to know where the vehicle is parked. It is a usual situation to park the vehicle in a parking lot and hours later when the driver intends to go back to look for its car, to forget exactly where the car is parked. VAiPho automatically stores and detects where the vehicle is parked in order to inform about it to the user, if required. The theoretical concept behind this algorithm is also simple. When the application detects that the vehicle is turned off, it automatically stores the GPS coordinates in the database so that if the user desires, it can access the application and show where the vehicle is parked.

•    Traffic Jam: This event has also an associated timeout so that it is not stored forever in the database. As in the free parking event, there is a parameter that is added to the traffic jam detection time that indicates whether the event is no longer valid, in which case, it is deleted from corresponding table. The jam detection is based on the idea that the type of road and travel direction of a vehicle determines its speed, so that when this is abnormally low, a possible traffic event is generated.

•    Data Aggregation: The aggregation module is aimed at testing matches in the traffic jam detection in real time, in order to avoid malicious announcements. When the jam event is detected, the program generates an instance of the class that handles the event. What this class does internally is to create an instance of another class 'Aggregation' and to start a cycle that aims to corroborate the veracity of the traffic jam detection event. The first step of this class is sending a new set of aggregation. This packet is sent to all users who are authenticated with the user that created it. The first packet of this type is the packet (I), which is generated from the detection of an event (T1). Once the packet (I) is managed, T1 is called the Information Aggregation event, what involves checking whether the event T1 is stored in the database. If there is no event in the database, the node checks that the coordinates of this traffic jam are within 100 meters radius of the coordinates where the user is. If so, it checks that the event occurs in the area so that if this checking process is positive, it looks for the event in the table of the events that have not yet been corroborated. If not found in this table, the event is first inserted in it, and then signed and sent in a packet of type T1 and sub-type F. On the other hand, if it is already in the table, that means that looking for the oldest data to delete them in order to avoid unnecessary duplication of events in the table. Finally, there is a packet T1 to be received by all users and to be added by them to the corresponding event table. There is another function responsible for verifying the authenticity of the signatures. If the signatures are valid, they are added to the events table and notified to VAiPho. Otherwise, the corresponding entry is removed from the possible events tables, and VAiPho broadcasts a notification. The other options to get a packet of an authenticated user are of P1 and P2 type, corresponding to advertising packets and free parking packets. These are dealt in the same way: decoding a received data and inserting them into the events table. This insertion is done so that it avoids unnecessary duplication.

• GPS Module: One of the most critical aspects of this work was concerning the geo-positioning technology because with Symbian we had many problems to graphically display what VAiPho can detect. Finally we opted for a provisional solution based on events painted on a coloured background.

• Bluetooth Module: One of the premises of VAiPho is to serve as an assistance tool for the driver. This requires a design that does not affect the attention of the driver on the road. To fulfil this, the events are shown on the map and at the same time they are announced through voice messages. Furthermore, both initiation and shutdown of the application are fully automatic. These features are possible in VAiPho thanks to Bluetooth technology. The used idea is simple, as it consists in starting and shutting down VAiPho when it automatically detects that the vehicle is started or shut down. In particular, in VAiPho implementation, since the free-hands device of the vehicle is started and turned off together with the car, we have taken advantage of it because the smartphone is able to connect to the car's Bluetooth, so the application starts with the free-hands device, which starts with the car.

## 4. INTERFACES OF THE PROPOSAL

VAiPho is formed by a series of interfaces that are used together to function as a single distributed application.

• VAiPho WatchDog: This interface, shown in Figure 2, is responsible that VAiPho program starts automatically. It runs in background and starts when the smartphone starts.

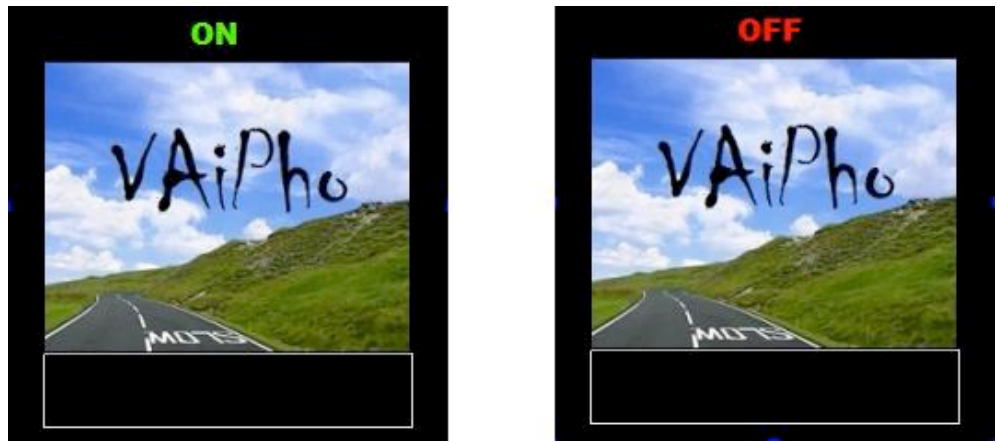

*Figure 2. Visual aspect of VAiPho WatchDog*

VAiPho User Interface: This interface allows the configuration of all VAiPho parameters and variables, as shown in Figure 3, as well as the use of the finder application. One of those variable parameters is the possibility that VAiPho turns off when the phone battery reaches a certain threshold, in order to preserve battery for the main function of the phone, which is to make or receive phone calls. This interface allows using VAiPho to know where the parked vehicle is. After pressing the button, the position where the parked vehicle is with respect to the user position appears on the map. Furthermore, this is the only interface that the user runs manually, since the other interface is automatically launched when the smartphone connects with the car´s Bluetooth.

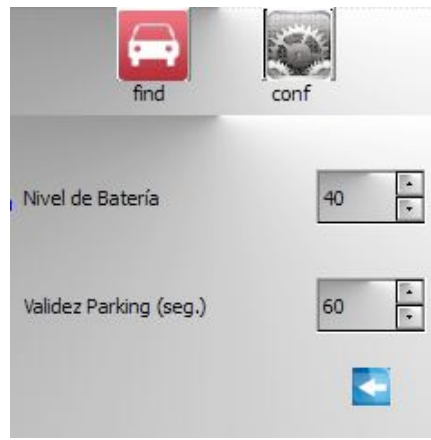

*Figure 3. Visual aspect of VAiPho User Interface*

• VAiPho Automatic: This interface, shown in Figure 4, can be considered the primary interface because it is an application that handles device authentication, user identification, event notification, communications and data aggregation.

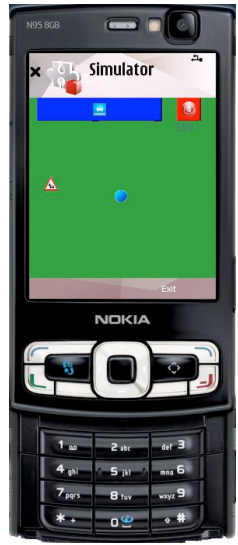

*Figure 4. Visual aspect of VAiPho Automatic*

Finally, Table 4 shows the power consumption involved in the use of different interfaces of VAiPho for Symbian. The worst case of battery life with different interfaces is there shown.

*Table 4. Smartphone's battery duration.*

| Interfaces | Time in hours |
|---|---|
| SmartPhone without VAiPho | 84 |
| SmartPhone with VAiPho WatchDog ON | 30 |
| SmartPhone with VAiPho Automatic ON | 4 |

## 5. CONCLUSIONS

In recent years, many measures have been taken by governments to try to manage the excessive number of vehicles on the roads, but without producing substantial results. Road safety is and will be a primary objective in society because every day there are more vehicles, so that consequently traffic accidents remain the third leading cause of death. The scientific community has been doing research on different solutions to enable the efficient management of traffic through the creation of vehicular networks that allow communication between vehicles, so that drivers can know real-time circumstances happening on the nearby roads. In order to implement the solution according to how it has been proposed in the literature, a high cost would be necessary both for government, which should install and maintain new communication infrastructure on roads, and for users, which would have to adapt or change their vehicles. By applying VAiPho, these economic investments can be avoided because it makes the VANET available today with existing technology such as a simple smartphone. In particular, this work has analyzed the case of the implementation of a mobile phone of the Nokia Company with the Symbian operating system. The communications of the network have been implemented in this work thanks to secure cryptographic algorithms included in VAiPho. In addition, the automatic event detection provides full confidence in its accuracy because it allows the detection of erroneous or malicious false messages. Thus, a tool for driver assistance has been born. In particular, the Symbian implementation of VAiPho has been developed to enable compatibility with other platforms like Maemo and Meego. Regarding future works, the battery consumption of VAiPho WatchDog is a key issue that needs improvement in future versions. On the other hand, the immense range of possible new applications that this work opens is really spectacular. Currently, VAiPho is created to detect traffic jams and free parkings, and to store the exact location where the vehicle is parked, while providing a geolocated advertising platform. However, in the future, many more features may be added in an easy and intuitive way to VAiPho for Symbian, thanks to the modular design that has been followed in its implementation.

## ACKNOWLEDGEMENTS

Research supported by the Ministerio de Ciencia e Innovación and the European FEDER Fund under Project TIN2011-25452, FPI scholarship BES-2009-016774, and ACIISI FPI scholarship BOC Number 60.

### *Information about the authors:*

**F. Martín-Fernández** – received his Computer Science Engineering Degree from the University of La Laguna. He is now starting with his research on security in VANETs and Internet of Things. He belongs to the CryptULL research group devoted to the development of projects on cryptology.

**C. Caballero-Gil** – received his Computer Science Engineering Degree from the University of Las Palmas de Gran Canaria (España) and his PhD from the University of La Laguna. His research is on VANET security. He belongs to the CryptULL research group devoted to the development of projects on cryptology, and is involved in several projects and publications related to this area. He has authored several conference and journal papers.

**P. Caballero-Gil** – graduated with a BSc and a PhD in Mathematics from the University of La Laguna in 1990 and 1995, respectively. Since 1990 she has been with the University of La Laguna at the Department of Statistics, Operations Research and Computation where she is Professor of Computer Science and Artificial Intelligence. Her area of expertise includes security of wireless networks, cryptanalysis and cryptographic protocols. She leads the CryptULL research group devoted to the development of projects on Cryptology. She has authored many refereed conference papers, journal articles and books.

**J. Molina-Gil** – received her Computer Science Engineering Degree from the University of Las Palmas de Gran Canaria (España) and her PhD from the University of La Laguna. Her research is on VANET security. She belongs to the CryptULL research group devoted to the development of projects on cryptology, and is involved in several projects and publications related to this area. She has authored several conference and journal papers.